\documentstyle[12pt,epsfig]{article}

\newcommand{\be}{\begin{equation}}
\newcommand{\ee}{\end{equation}}
\newcommand{\Dlt}{\Delta}
\newcommand{\dlt}{\delta}
\newcommand{\br}{{\bf r}}
\newcommand{\ba}{{\bf a}}

\newcommand{\al}{\alpha}

\newcommand{\gm}{\gamma}
\newcommand{\om}{\omega}
\newcommand{\Om}{\Omega}
\newcommand{\dgr}{\dagger}

\begin{document}

\begin{center}

{\Large {\bf Regulating atomic imbalance in double-well
lattices} \\ [5mm]

V.I. Yukalov$^1$ and E.P. Yukalova$^2$} \\ [3mm]

{\it
$^1$Bogolubov Laboratory of Theoretical Physics, \\
Joint Institute for Nuclear Research, Dubna 141980, Russia \\ [3mm]

$^2$Department of Computational Physics, Laboratory of Information
Technologies,\\
Joint Institute for Nuclear Research, Dubna 141980, Russia}

\end{center}

\begin{abstract}

An insulating optical lattice with double-well sites is considered.
In the case of the unity filling factor, an effective Hamiltonian in
the pseudospin representation is derived. A method is suggested for
manipulating the properties of the system by varying the shape of
the double-well potential. In particular, it is shown that the atomic
imbalance can be varied at will and a kind of the Morse-alphabet sequences
can be created.

\end{abstract}

\vskip 1cm

{\parindent=0pt
{\bf PACS}: 03.75.Lm; 03.67.Hk; 03.67.Lx 

\vskip 1cm

 }

\newpage

\section{Introduction}

Physics of cold atomic gases, Bose [1--8] and Fermi [9,10], is a fastly 
developing field of research, both theoretically and experimentally.
Cold atoms in optical lattices provide a versatile tool for various
applications [11--14]. Recently, a novel type of optical lattices has 
been realized in experiments [15--17], the so-called double-well lattices. 
Each site of such a lattice is formed by a double-well potential.

Generally, the theoretical description of the double-well lattices is
essentially more complicated than that of the usual lattices. Therefore,
in order to describe the properties of such lattices, it is useful to
consider some particular types of them.

In the present paper, we consider a particular kind of the double-well
lattice, which is characterized by the following main features: atoms in 
the lattice are in an insulating state, and the atomic filling factor is 
equal to unity. Such a type of a lattice is of principal importance 
providing a convenient setup for realizing quantum information processing 
and quantum computing [11]. Another important property of the lattices, 
we shall be considering, is the existence of atomic interactions between 
different lattice sites. The principal goal of the present paper is to 
demonstrate that the atomic imbalance between the wells of each double 
well can be manipulated and that arbitrary sequences of the order-disorder 
transition can be generated.

\section{Insulating double-well lattice}

We start with the standard form of the energy Hamiltonian, expressed
through the field operators in the Heisenberg representation. Since the
system is assumed to be in the insulating state, the field operators can
be expanded over localized orbitals, which yields the Hamiltonian
\be
\label{1}
\hat H = \sum_{nj} E_n c_{nj}^\dgr c_{nj} \; + \;
\frac{1}{2} \; \sum_{ \{ j\} } \; \sum_{ \{ n\} }
\Phi_{j_1j_2j_3j_4}^{n_1n_2n_3n_4} c_{n_1j_1}^\dgr  c_{n_2j_2}^\dgr
c_{n_3j_3} c_{n_4j_4} \; ,
\ee
in which $c_{nj}$ is a field operator labelled by the quantum index $n$
and the site number $j$, related to a lattice vector $\ba_j$. The atoms
can be either bosons or fermions, with the operator commutation relations
$$
\left [ c_{mi},\; c_{nj}^\dgr \right ]_\mp \; = \; 
\dlt_{mn} \dlt_{ij} \; , \qquad [ c_{mi},\; c_{nj} ]_\mp \; = \; 0 \; ,
$$ where the commutator is assumed for bosons and anticommutator for 
fermions. The value $E_n$ represents the energy levels of an atom in 
a double-well located at a site of the lattice. The quantity 
$\Phi_{j_1j_2j_3j_4}^{n_1n_2n_3n_4}$ is a matrix element of the 
interaction potential with respect to the localized orbitals labelled 
by the indices $n$ and $j$. We assume that the interaction potential is 
sufficiently strong, so that the interactions of atoms between at least 
nearest-neighbor sites cannot be neglected. This can be easily achieved, 
for instance, with the long-range potentials, such that exist between 
polar molecules and between Rydberg atoms [18] or between the atoms with 
large magnetic moments [19]. A known example of the latter atoms is Cr 
that can be cooled to ultracold temperatures [20]. In the case of ions, 
long-range forces would be due to the Coulomb interaction.

The assumption that each lattice site contains just a single atom can be
formalized by means of the unipolarity condition
\be
\label{2}
\sum_n c_{nj}^\dgr c_{nj} = 1 \; , \qquad c_{nj} c_{nj} = 0 \; .
\ee
This is the known condition, used earlier by Bogolubov [21] for treating 
ferromagnets. Keeping in mind low temperatures, we can consider only two 
lowest energy levels of a double-well potential, enumerated by $n=1,2$, 
so that $E_1<E_2$. As is known [22], the ground-state orbital, of an atom 
in a double-well potential, is symmetric with respect to spatial inversion,
while that for the first excited state is antisymmetric.

Retaining two lowest levels allows us to invoke the pseudospin
representation by introducing the pseudospin operators
$$
S_j^x = \frac{1}{2} \left ( c_{1j}^\dgr c_{1j} -
c_{2j}^\dgr c_{2j} \right ) \; ,
$$
\be
\label{3}
S_j^y = \frac{i}{2} \left ( c_{1j}^\dgr c_{2j} -
c_{2j}^\dgr c_{1j} \right ) \; , \qquad
S_j^z = \frac{1}{2} \left ( c_{1j}^\dgr c_{2j} +
c_{2j}^\dgr c_{1j} \right ) \; .
\ee
To understand the meaning of these operators, one can introduce the left 
and right location operators, respectively, as
\be
\label{4}
c_{jL} \equiv \frac{1}{\sqrt{2}}\; ( c_{1j} + c_{2j} ) \; , \qquad
c_{jR} \equiv \frac{1}{\sqrt{2}}\; ( c_{1j} - c_{2j} ) \; .
\ee
Then operators (3) become
$$
S_j^x = \frac{1}{2} \left ( c_{jL}^\dgr c_{jR} + c_{jR}^\dgr c_{jL}
\right ) \; , \qquad S_j^y = -\; \frac{i}{2} \left ( c_{jL}^\dgr c_{jR} -
c_{jR}^\dgr c_{jL} \right ) \; ,
$$
\be
\label{5}
S_j^z = \frac{1}{2} \left ( c_{jL}^\dgr c_{jL} - c_{jR}^\dgr c_{jR}
\right ) \; .
\ee
Note that this representation is valid for both statistics, so that atoms
can be either bosons or fermions. The physical meaning of these operators
is as follows: $S_j^x$ characterizes the tunneling intensity between the
wells of a double-well potentials, $S_j^y$ describes the Josephson current
between the wells, and $S_j^z$ defines the atomic imbalance between the
wells. It is worth emphasizing that for a double well it is necessary to
take into account at least two lowest energy levels, but not just one
ground state, since only with two levels there exists tunneling between
the wells. Actually, it is exactly because of the tunneling that the energy
levels split into pairs [22].

Let us use the notation 
$$
E_0\equiv \frac{E_1+E_2}{2} \; ,
$$ 
and
$$
V_{ij}^{mnm'n'}\equiv\Phi_{ijji}^{mnm'n'}\pm\Phi_{ijij}^{mnn'm'} \; ,
$$
where the sign plus or minus is for bosons or fermions, respectively. Also, 
let us denote the matrix elements
$$
A_{ij} \equiv \frac{1}{4} \left ( V_{ij}^{1111} + V_{ij}^{2222} +
2 V_{ij}^{1221} \right ) \; , \qquad
B_{ij} \equiv \frac{1}{2} \left ( V_{ij}^{1111} + V_{ij}^{2222}-
2 V_{ij}^{1221} \right ) \; ,
$$
\be
\label{6}
C_{ij} \equiv \frac{1}{2} \left (  V_{ij}^{2222} - V_{ij}^{1111}
 \right ) \; , \qquad I_{ij} \equiv - 2 V_{ij}^{1122} \; .
\ee
An important quantity is the tunneling frequency
\be
\label{7}
\Om \equiv E_2 - E_1 + \sum_{j(\neq i)} C_{ij}
\ee
characterizing the atomic tunneling between the wells of a double-well
potential.

Using the above conditions and notations, we reduce Hamiltonian (1) to the
pseudospin form
\be
\label{8}
\hat H = E_0 N \; + \; \frac{1}{2} \; \sum_{i\neq j} A_{ij} \; - \;
\Om \sum_j S_j^x  \; + \; \sum_{i\neq j} B_{ij} S_i^x S_j^x \;
- \; \sum_{i\neq j} I_{ij} S_i^z S_j^z \; .
\ee
The first two terms do not contain operators, so do not play role in what
follows. The magnitude of $B_{ij}$ can be comparable with $\Om$, hence,
it cannot be neglected. The value of the tunneling frequency $\Om$ can be
varied in a wide range, depending on the shape of the double-well potential.
To illustrate this, let us take the latter in the form
$$
V(\br) = V_0 \left ( \frac{r_x}{r_0} \right )^2 \left [
\left ( \frac{r_x}{r_0} \right )^2 -2 \right ] + V_{yz} \; ,
$$
where $V_{yz}$ is, say, a harmonic potential in the $y$- and $z$-directions.
Then the tunneling frequency strongly depends on the parameter
\be
\label{9}
\al \equiv \frac{1}{\sqrt{2mr_0^2 V_0^2} } =
\frac{a}{\pi r_0} \; \sqrt{\frac{E_R}{V_0} } \; ,
\ee
in which $a$ is the lattice spacing in the $x$-direction and
$E_R\equiv\pi^2/2ma^2$ is the recoil energy. By varying the interwell
distance $r_0$ or the potential depth $V_0$, the parameter $\al$ can be
varied in a wide range between $\al\ll 1$ and $\al\gg 1$. Direct calculations
[22] show that the tunneling frequency $\Om$ can also be made either small or
large, such that
$$
\Om \simeq 6V_0 \exp\left ( - 2\pi\; \frac{r_0}{a} \;
\sqrt{\frac{V_0}{E_R} } \right ) \qquad (\al \ll 1) \; ,
$$
\be
\label{10}
\Om \simeq 9V_0 \left ( \frac{a}{\pi r_0} \right )^{2/3} \;
\left (\frac{E_R}{V_0} \right )^{1/3}\qquad (\al \gg 1) \; .
\ee
That is, $\Om$ can be regulated by governing the shape of the double-well
potential. And varying $\Om$ it is possible to regulate the system properties.

The important system characteristics are the average tunneling intensity
\be
\label{11}
x \equiv \frac{2}{N_L} \; \sum_j < S_j^x > \; ,
\ee
average Josephson current
\be
\label{12}
y \equiv \frac{2}{N_L} \; \sum_j < S_j^y > \; ,
\ee
and the average well imbalance
\be
\label{13}
z \equiv \frac{2}{N_L} \; \sum_j < S_j^z > \; ,
\ee
where $N_L$ is the number of lattice sites and $<\cdot>$ implies statistical
averaging. These values depend on the system parameters, such as the
dimensionless transverse interaction
\be
\label{14}
b \equiv \frac{B}{I+B} \; ,
\ee
in which
$$
B \equiv \frac{1}{N_L} \; \sum_{i\neq j} B_{ij} \; , \qquad
I \equiv \frac{1}{N_L} \; \sum_{i\neq j} I_{ij} \; ,
$$
and the dimensionless tunneling frequency
\be
\label{15}
\om \equiv \frac{\Om}{I+B} \; .
\ee

\section{Manipulating atomic imbalance}

The equations of motion for the average quantities (11) to (13) can be obtained
by averaging the Heisenberg equations for the spin operators $S_j^\al$. In this
procedure, we assume that the system is at zero temperature, we employ the
local-field approximation [23], taking into account particle interactions
occurring in the local field of other particles. Measuring time in units of
$1/(I+B)$, we obtain
$$
\frac{dx}{dt} = ( 1 - b) yz - \gm_2 ( x - x_t) \; , \qquad
\frac{dy}{dt} = (\om-x) z - \gm_2 (y - y_t) \; ,
$$
\be
\label{16}
\frac{dz}{dt} = (bx-\om) y - \gm_1 ( z - z_t)  \; ,
\ee
where $\gm_1$ and $\gm_2$ are the attenuation parameters that are expressed
through atomic interactions $I_{ij}$ and $B_{ij}$ similarly to the corresponding
damping parameters in spin systems [24], and where the local fields are
\be
\label{17}
x_t = \frac{\om-bx}{h} \; , \qquad y_t = 0 \; , \qquad z_t =
\frac{1-b}{h}\; z \; ,
\ee
with 
$$
h=\sqrt{(\om-bx)^2+(1-b)^2z^2} \; .
$$ 
By their form, the local fields (17) correspond to the equilibrium solutions 
for averages (11) to (13). However, since here $x=x(t)$ and $z=z(t)$ depend on 
time, the local fields (17) describe the locally equilibrium state.

Accomplishing the Lyapunov stability analysis for the system of equations (16),
we find two fixed points. One is given by the equations
\be
\label{18}
x_1^* = \om \; , \qquad y_1^*=0 \; , \qquad z_1^* =\sqrt{1-\om^2} \qquad
(\om < 1) \; ,
\ee
this fixed point being stable for $\om<1$, but unstable for $\om>1$. And the
other fixed point is
\be
\label{19}
x_2^* = 1 \; , \qquad y_2^*=0 \; , \qquad z_2^* = 0 \qquad
(\om > 1) \; ,
\ee
which is stable for $\om>1$, but unstable for $\om<1$.

It is worth emphasizing that the fixed points (18) and (19) are stable, for
the corresponding values of $\om$, only when $\gm_1$ and $\gm_2$ are not zero.
When the latter parameters are zero, the dynamical system (16) is structurally
unstable. This stresses the necessity of taking into account the attenuation
effects, without which there would be no correct description of dynamics.

The stationary solution (18) characterizes the phase of the system with a
nonzero well imbalance $z_1^*>0$. Hence, this can be called the ordered phase.
The stationary solution (19) describes a disordered phase, where the well
imbalance is zero, $z_2^*=0$. In an equilibrium system, the transition between
the ordered and disordered phases would correspond to a quantum phase 
transition, with the well imbalance playing the role of an order parameter. 
For a nonequilibrium system, the value $\om=1$ is a bifurcation point, where 
a dynamical phase transition takes place.

As has been discussed above, the magnitude of the tunneling frequency can be
varied in a wide range by changing the shape of the double well. This means that
we have a straightforward opportunity of regulating the state of the double-well
lattice by reswitching the value of $\om$ between that one corresponding to the
ordered state and another related to the disordered state. That is, we can
manipulate with the well imbalance, reswitching it, according to our will,
between the zero and nonzero values.

As an illustration of this remarkable feasibility of regulating the well
imbalance, we solve numerically Eqs. (16), with a time-modulated tunneling
frequency $\om=\om(t)$, which is varied according to the rule
\begin{eqnarray}
\label{20}
\om(t) = \left \{ \begin{array}{ll}
\om_1 , & \; 0 \; \leq \; t \; < \; \Dlt t_1 \\
\\
\om_2, & \; \Dlt t_1 \; \leq t \; < \; \Dlt t_1 + \Dlt t_2 \\
\\
\om_1, & \; \Dlt t_1 + \Dlt t_2 \; \leq t \; < \;
\Dlt t_1 + \Dlt t_2 + \Dlt t_3 \; , \\
\ldots & \ldots\ldots\ldots
\end{array} \right.
\end{eqnarray}
where $\om_1<1$ and $\om_2>1$. The time intervals $\Dlt t_n$ can be chosen
arbitrarily. For instance, we can take all of them being equal, as in Fig.
1, where $\Dlt t_1=\Dlt t_2=\ldots=\Dlt t_n$. Or we can organize a periodic
sequence with two unequal time intervals $\Dlt t_1=\Dlt t_3=\ldots=\Dlt
t_{2n+1}$ and $\Dlt t_2=\Dlt t_4=\ldots=\Dlt t_{2n}$, as in Fig. 2. Finally,
we can take arbitrary time intervals $\Dlt t_n$,  as in Fig. 3, realizing the
sequences, according to our will, in which information, as in the Morse
alphabet, could be encoded. Such Morse-alphabet sequences for the pseudospin
system, representating a double-well lattice, are analogous to the punctuated
sequences for superradiating spin systems [25].

In conclusion, we have derived a pseudospin representation for insulating
double-well lattices, with the unity filling factor. Such a system can be
either in an ordered or in a disordered state, which are characterized by a
nonzero or zero well imbalance, respectively. For correctly describing the
system dynamics, one has to take account of relaxation effects, without which
the related dynamical system is structurally unstable. By a temporal modulation
of the lattice parameters, e.g., of the tunneling frequency, it is possible to
regulate the well imbalance, organizing the Morse-alphabet sequences, which can
be employed for quantum information processing. The possibility of regulating 
the well imbalance in a double-well lattice, by organizing arbitrary temporal 
sequences of ordered and disordered states, is the main result of the present 
paper.

\vskip 5mm

{\bf Acknowledgement}

\vskip 2mm

Financial support from the Russian Foundation for Basic Research
(Grant 08-02-00118) is appreciated.

\newpage

\begin{center}
{\large{\bf Figure Captions}}

\end{center}

\vskip 5mm

{\bf Fig. 1}. Population imbalance as a function of dimensionless time
for the periodic reswitching of the tunneling between $\om_1=0.1$ and
$\om_2=5$, with equal time intervals $\Dlt t_1 = \Dlt t_2=10$. The system
parameters are $\gm_1=\gm_2=1$ and $b=0.5$. Initial conditions are
$x_0=0.33$, $y_0=0.8$, $z_0=0.5$.

\vskip 5mm

{\bf Fig. 2}. Population imbalance as a function of dimensionless time
for the periodic reswitching of the tunneling between $\om_1=0.1$ and
$\om_2=1.1$, with unequal time intervals $\Dlt t_1=10$ and $\Dlt t_2=25$.
Other parameters are as in Fig. 1.

\vskip 5mm

{\bf Fig. 3}. Population imbalance as a function of dimensionless time
for a nonperiodic reswitching of the tunneling between $\om_1=0.1$ and
$\om_2=1.1$, with varying time intervals $\Dlt t_j$. Other parameters are
as in Fig. 1.

\newpage

\begin{figure}[ht]
\centerline{\psfig{file=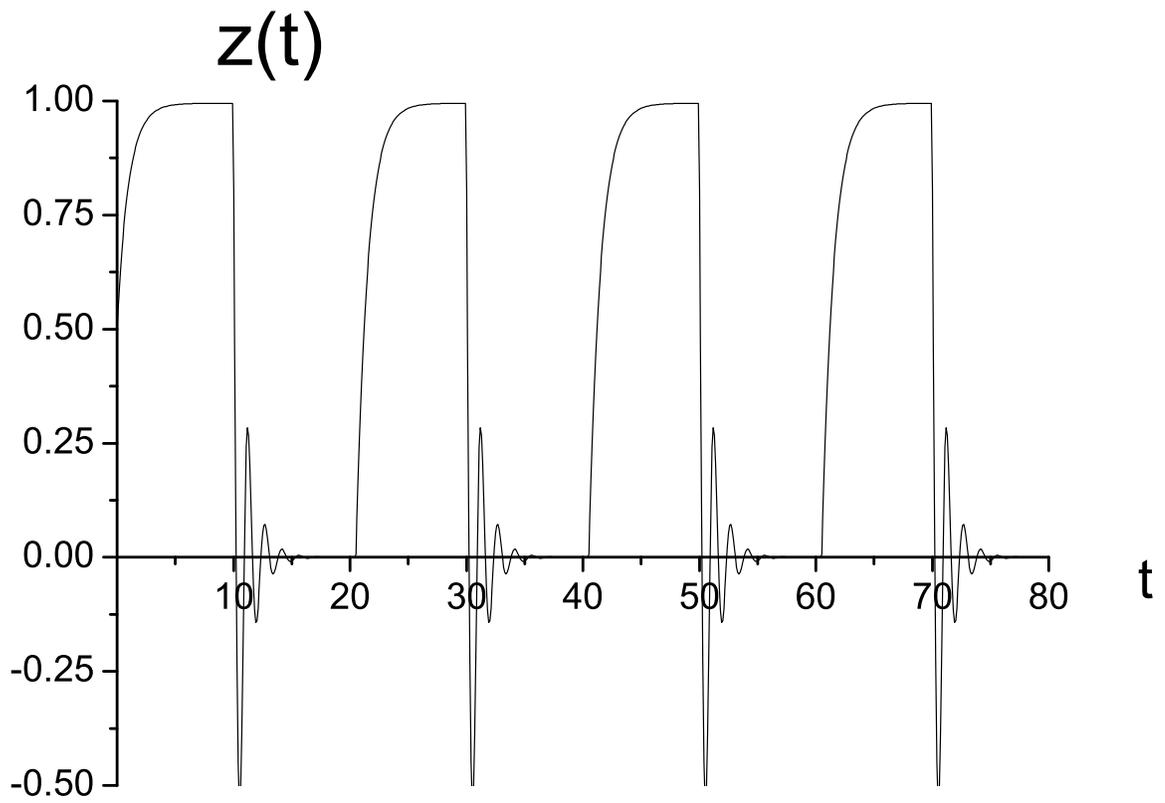,height=5in}}
\caption{Population imbalance as a function of dimensionless time
for the periodic reswitching of the tunneling between $\om_1=0.1$ and
$\om_2=5$, with equal time intervals $\Dlt t_1 = \Dlt t_2=10$. The system
parameters are $\gm_1=\gm_2=1$ and $b=0.5$. Initial conditions are
$x_0=0.33$, $y_0=0.8$, $z_0=0.5$.}
\label{fig:Fig.1}
\end{figure}

\newpage

\begin{figure}[ht]
\centerline{\psfig{file=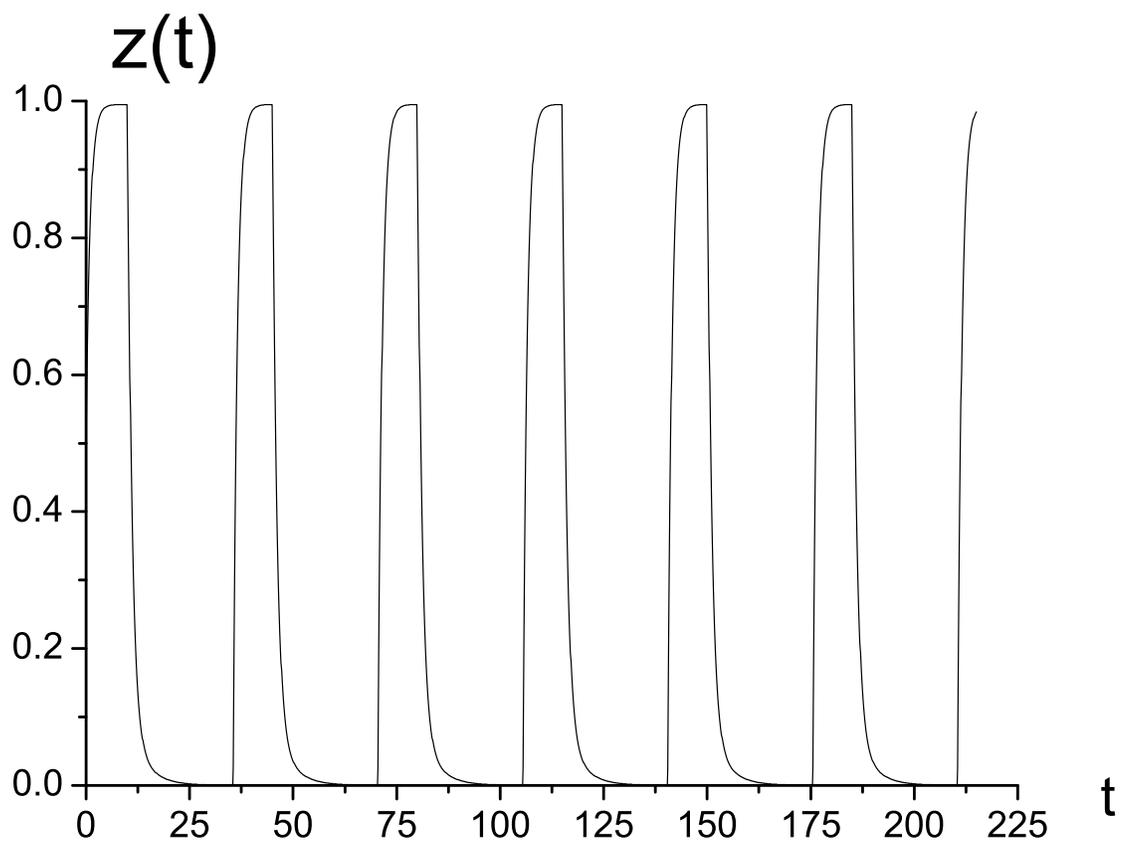,height=5in}}
\caption{Population imbalance as a function of dimensionless time
for the periodic reswitching of the tunneling between $\om_1=0.1$ and
$\om_2=1.1$, with unequal time intervals $\Dlt t_1=10$ and $\Dlt t_2=25$.
Other parameters are as in Fig. 1.}
\label{fig:Fig.2}
\end{figure}

\newpage

\begin{figure}[ht]
\centerline{\psfig{file=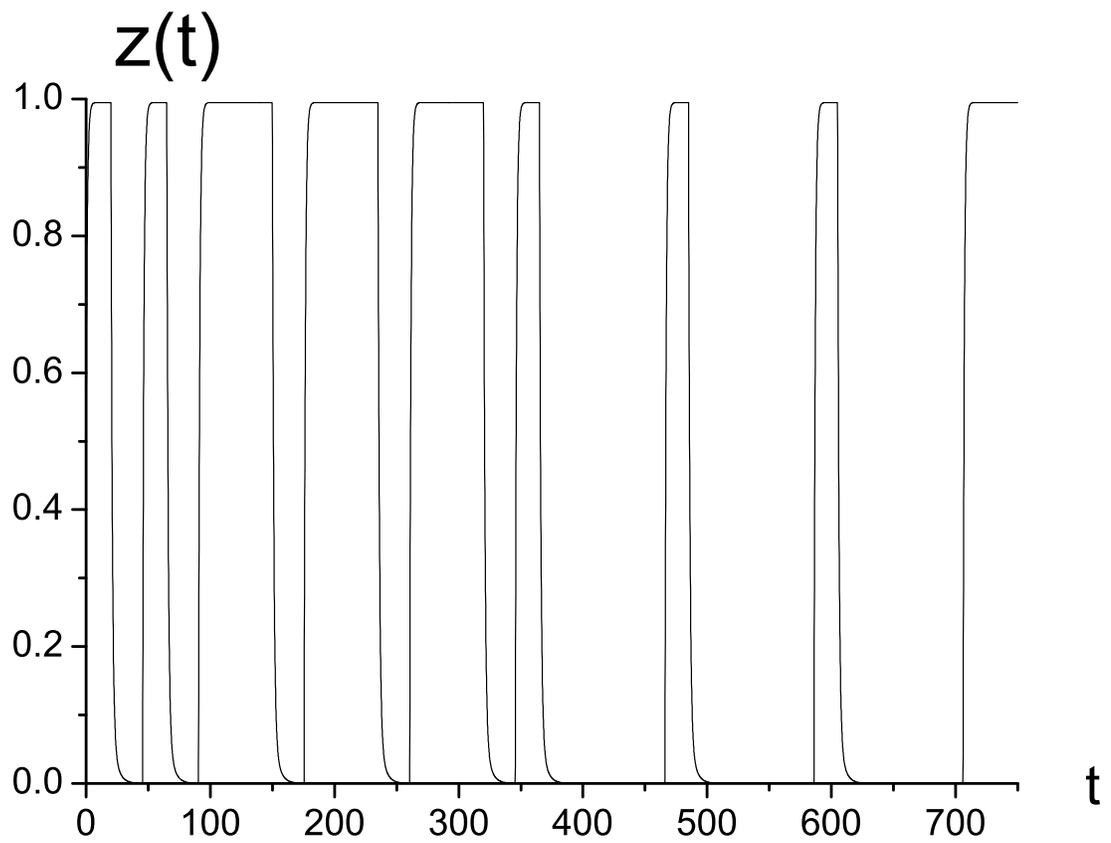,height=5in}}
\caption{Population imbalance as a function of dimensionless time
for a nonperiodic reswitching of the tunneling between $\om_1=0.1$ and
$\om_2=1.1$, with varying time intervals $\Dlt t_j$. Other parameters are
as in Fig. 1.}
\label{fig:Fig.3}
\end{figure}

\end{document}